\documentclass[twocolumn,prl,showpacs,superscriptaddress]{revtex4}
\usepackage{epsf}
\usepackage{color}
\usepackage{verbatim}
\usepackage{hyperref}
\usepackage{graphics}
\usepackage{graphicx}
\usepackage{natbib}
\usepackage{dcolumn}
\usepackage{bm}

\newcommand {\nigas}{NiGa$_2$S$_4$ }
\newcommand {\nigasn}{NiGa$_2$S$_4$}

\begin{document}
\title{Halperin-Saslow modes as the origin of the low temperature anomaly in NiGa$_2$S$_4$}
\author{Daniel Podolsky}
\affiliation{Department of Physics, University of Toronto, Toronto,
Ontario M5S 1A7 Canada}
\author{Yong Baek Kim}
\affiliation{Department of Physics, University of Toronto, Toronto,
Ontario M5S 1A7 Canada}
\affiliation{School of Physics, Korea Institute for Advanced Study, Seoul 130-722, Korea }
\date{\today}
\begin{abstract}
The absence of magnetic long range order in the triangular lattice spin-1 antiferromagnet 
\nigas \cite{nakatsuji1} has 
prompted the search for a novel quantum ground state. In particular, several
experiments suggest the presence of a linearly dispersing mode despite no long-range 
magnetic order \cite{nakatsuji1,nakatsuji2,nakatsuji3,stock}. 
We show that the anomalous low temperature properties of \nigas can naturally be explained
by the formulation developed by Halperin and Saslow\cite{HalperinSaslow} where the linearly dispersing
Halperin-Saslow mode may exist in the background of frozen spin moments and zero net magnetization.
We provide highly non-trivial consistency checks on the existing experimental data and suggest future
experiments that can further confirm the existence of the Halperin-Saslow mode.  Our results place strong 
constraints on any microscopic theory of this material.
\end{abstract}
\maketitle

{\it Introduction.}-- Frustrated magnets are an excellent playground for the discovery of novel
quantum phases of matter.  The intricate interplay between geometrical frustration and quantum fluctuations lies at the heart of the physical mechanism for such possibilities. Recently, various frustrated magnets with small spin ($S=\frac{1}{2}$ and $S=1$) have been 
discovered \cite{nakatsuji1,recent_exp}.  
These experiments provide great opportunities for studying quantum as well as thermal fluctuation effects in frustrated magnets. In particular, \nigas is an insulator composed of spin-1 Ni atoms on a triangular lattice,
that shows no long range magnetic order in neutron scattering down to at least 0.35 K, despite an antiferromagnetic Curie-Weiss temperature $|\Theta_{CW}|= 80$ K \cite{nakatsuji1}.
Instead, spin freezing is observed at $T_f= 8$ K, a temperature that also serves as the onset for a variety of unusual experimental signatures.

Below $T_f$, the spin susceptibility saturates to a constant value, while the magnetic part of the specific heat acquires a $T^2$ power law \cite{nakatsuji1}.  This $T^2$ dependence of the specific heat is consistent with the presence of gapless hydrodynamic modes propagating in two dimensions, provided that these modes are coherent over a length scale $L_0$ that exceeds 500 lattice spacings \cite{nakatsuji1}.  On the other hand, elastic neutron scattering  on powder \cite{nakatsuji1} and single crystal \cite{stock} samples show no evidence of long range magnetic order.  Instead, coincident with the spin freezing temperature,  there is the onset of {\it static, yet short range} 
magnetic correlations, with a correlation lenght $\xi$ of order of only 7 lattice spacings. In addition, nuclear magnetic resonance (NMR) measurements have
observed a spin-lattice relaxation rate $1/T_1 \propto T^3$ for $T\ll T_f$  \cite{takeya}.   Recent inelastic neutron scattering measurements observe a linearly dispersing excitation mode \cite{stock}

In order to account for some of these results, there have been recent proposals of the formation of subtle types of long-range spin ordering.   Based on models with
large biquadratic interactions, Tsutentsugu and Arikawa
proposed a three-sublattice nematic state \cite{tsunetsugu} and
Bhattacharje {\it et al.}~proposed a uniform nematic state \cite{senthil} 
(see also Ref.\cite{mila} for related theoretical models).
On the other hand, Kawamura and Yamamoto proposed a state with 
bound $Z_2$ vortices \cite{kawamura}.
The nematic states of Refs.~\cite{tsunetsugu} and \cite{senthil} may be consistent with the constant spin susceptibility, and also contain gapless director modes, which can account for the $T^2$ specific heat. 
However,  these proposals leave a few questions unanswered.  For instance, there is no clear connection between the onset of nematic order and the appearance of static short-range magnetic correlations seen at $T_f$.  In addition, a nematic state does not account for the linearly dispersing inelastic neutron scattering peak, or for the $1/T_1\sim T^3$ spin-lattice relaxation rate -- the director modes of a nematic do not couple directly either to the neutrons or to the atomic spins probed by NMR\cite{mila}. 

In this Letter, we will argue that these experimental discoveries can be 
explained by the presence of gapless Halperin-Saslow (HS) modes \cite{HalperinSaslow}.  
Halperin and Saslow originally proposed the existence of hydrodynamic modes in the context of spin-glasses.  However, 
as discussed below, HS modes can exist in a much broader class of materials.   HS modes have a linear dispersion, and they couple directly to atomic spins and also to neutrons.  Therefore, they can account for the full phenomenology of \nigas in a natural way.  In the rest of the paper, we will provide various qualitative and semi-quantitative consistency checks, and also discuss future experimental consequences of the HS scenario.  
We are not aware of any confirmed example of the Halperin-Saslow modes, therefore
\nigas may present the first clear case for these long-sought low energy modes.   
Our discussion will be phenomenological in nature, as it will not address the microscopic mechanism by which the system chooses the HS state, but will rather rely on general hydrodynamic considerations.  This has the advantage that our results are not limited to \nigasn, but are easily adapted to other frustrated magnets that may be in the HS state. 

{\it Overview of the Halperin-Saslow theory.}--
Halperin and Saslow showed that a system with static moments can support low energy hydrodynamic modes, even in the
absence of periodic long range magnetic order \cite{HalperinSaslow}. Their
formalism is very general and relies on two assumptions only: the presence of static moments in  the (possibly metastable) 
ground state, and a finite spin stiffness to slow spatial deformations of the static spin texture.  If the moments are non-collinear and have zero net magnetization, the HS theory predicts the existence of three linearly-dispersing hydrodynamic modes.  The elastic neutron scattering experiments in \nigas indicate that there are static helical spin textures at temperatures below $T_f$.  The fact that these spin textures are short ranged is inconsequential to the HS scenario, provided that a finite spin stiffness survives at long distances.

For a system satisfying these assumptions, the free energy of a weakly-perturbed state above the ground state is \cite{HalperinSaslow}
\begin{eqnarray}
\Delta F[{\bf m},{\bf \theta}]=\frac{d}{2}\sum_{\alpha=1}^3\int d^2 r \left [ m_\alpha^2 \chi^{-1}+\rho_{s} (\nabla \theta_\alpha)^2\right ] .\label{eq:HSfreeEnergy}
\end{eqnarray}
Here $\chi$ is the spin susceptibility, $\rho_s$  is the spin stiffness,  $m^\alpha(r)$ is the local magnetization density, 
and $\theta_\alpha$ are locally-defined rotation angles that describe the deformation in the excited-state spin configuration.  Here $m^{\alpha}$ and $\theta_{\alpha}$ are canonically conjugate variables.
In Eq.~(\ref{eq:HSfreeEnergy}) we
have assumed that the system is composed of weakly coupled layers of thickness $d$.   

Starting from Eq.~(\ref{eq:HSfreeEnergy}) one obtains three degenerate polarizations of spin waves with dispersion $\omega_k = v k$, with velocity 
\begin{eqnarray}
v=\gamma\sqrt{\frac{\rho_s}{\chi}},
\end{eqnarray}
where $\gamma=g\mu_B/\hbar$ is the gyromagnetic ratio.
In a more general case, $\chi$ and $\rho_s$ are tensors, leading to three non-degenerate spin waves, with polarization-dependent velocities $v_j=\gamma(\rho_j/\chi_j)^{1/2}$.

{\it Basic consideration: contrasting specific heat and elastic neutron scattering data.}--
The presence of gapless HS modes is consistent with the  $T^2$ specific heat in \nigasn.  For linearly dispersing modes in two dimensions,
\begin{eqnarray}
\frac{c_M}{N_A \nu}=\frac{3\xi(3)}{\pi d}\frac{k_B^3 T^2}{\hbar^2}\sum_j \frac{1}{v_j^2}-\frac{3 k_B \pi}{L_0^2 d}.
\label{specific_heat}
\end{eqnarray}
Here, $\xi(3)=1.202\ldots$, $N_A$ is the Avogadro's number, the sum runs over polarizations $j$, $v_j$ is the collective mode velocity, and $\nu=\sqrt{3} a^2 d/2$ is the volume of a unit cell where we have 
assumed decoupled layers of thickness $d$ and $a$ is the in-plane lattice spacing.   
Here, we have allowed for the possibility that the collective modes are coherent only up to some length scale $L_0$.
Experimental fits yield a lower bound, $L_0>500 a$ \cite{nakatsuji1}.

Elastic neutron scattering measurements  on both powder and single crystal samples show that the freezing temperature $T_f$ coincides 
with the onset of static short-range-ordered moments.  Peaks appear at an incommensurate wave vector 
${\bf Q}=(1/6-\epsilon,1/6-\epsilon)$ .  The peaks are static, but they are not wave vector resolution-limited, corresponding to short-range 
order over a length $\xi_{xy}\sim 7 a$ in-plane, and $\xi_z\sim 6 \AA$ out-of-plane.  The wave vector ${\bf Q}$ is near the half-way point between the center and 
the corners of the Brillouine zone, and the spin moments seem to be arranged in a helical pattern.  This wave vector and spin structure 
would be expected from a system with a weak
ferromagnetic nearest neighbor coupling and a large antiferromangetic third-nearest neighbor coupling, $J_3\gg -
J_1>0$.  Such a hierarchy of couplings is consistent with semi-empirical cluster calculations \cite{takubo}, and a large $J_3>0$ is
also supported by {\it ab initio} calculations \cite{mazin}.

Note that the correlation length $\xi_{xy}$ is much smaller than the scale $L_0$, {\it i.e.} the collective modes responsible for the $T^2$ specific heat exist over a much larger length scale than the helical regions seen in neutron scattering.
Here we explore the possibility that below $T_f$, the system is in a quantum spin-glass-like state composed of helical domains that do not order at long distances, yet have a finite spin-stiffness.   One possible picture is to coarse grain the system, and to think of each helical domain as some block degree of freedom.  The residual interaction between blocks, of order $T_f$, is frustrated and disordered, leading to freezing behavior at $T_f$ in the absence of long-range helical order.  However, independently of the microscopic mechanism by which such a state may be formed, we focus on the generic properties that follow from the hydrodynamics of such a system.

{\it Two energy scales from the susceptibility and specific heat.}--
As a first semi-quantitative check, we would like to know whether the magnitude of the low temperature susceptibility and specific heat anomaly are consistent with the HS scenario.   In this scenario, these two quantities are related through the spin stiffness $\rho_s$, since $c_M/T^2 \propto 1/v^2\propto \chi_M/\rho_s$.   From $\chi_M$ and $c_M/T^2$, we can extract two energy scales,
\begin{eqnarray}
E_1&=&\frac{2 g^2\mu_B^2 S(S+1)}{z \chi_M / N_A}=k_B(113\,{\rm K}) \\
E_2&=&\frac{3 n_p \xi(3)}{\pi}\frac{k_B^3}{g^2\mu_B^2}\frac{\chi_M}{c_M/T^2}=k_B(6.5\,{\rm K}) 
\end{eqnarray}
where  $z=6$ is the coordination number of the lattice, and $n_p=3$ is the number of degenerate hydrodynamic modes.  Above, we have used the experimentally-measured values, $\chi_M(T\to 0)=0.0089$ emu/mole and $c_M/T^2=2.6\times 10^{-2}$ J mole$^{-1}$ K$^{-3}$ \cite{nakatsuji1}.   

The energy $E_1$ is related to the effective interaction between a spin and its environment, whereas $E_2$ corresponds to the spin
stiffness extracted from measurements.   Note that, while $E_1$ is comparable to the microscopic energy scale  $|\Theta_{\rm CW}|= 80$ K, $E_2$ is much smaller.  This is consistent with the HS scenario: the stiffness is expected to be considerably renormalized relative to the microscopic energies, to a scale of order of the freezing temperature $T_f=8$ K.

We can provide a theoretical upper bound on the spin stiffness.  This is done by computing the spin stiffness of
a long-range-ordered helical state, ignoring the fact that spin correlations in \nigas are short ranged.  Carrying out an analysis similar to Ref.~\cite{HalperinSaslow}, 
\begin{eqnarray}
\rho_{\alpha\beta}^{max}=- \frac{1}{2V}\left(\delta_{\alpha\beta}+\delta_{\alpha,3}\delta_{\beta,3}\right) \sum_{\vec{\mu}} J_{\vec{\mu}} \mu_x^2\sum_i \langle {\bf S}_i\cdot {\bf S}_{i+\vec{\mu}}\rangle \label{eq:helicalRho}
\end{eqnarray}
where the sum runs over all vectors $\vec{\mu}$ connecting lattice sites, $\mu_x=\vec{\mu}\cdot\hat{x}$ for an arbitrary direction $\hat{x}$, and $V$ is the volume of the sample.  Equation (\ref{eq:helicalRho}) is an upper bound on the stiffness of \nigasn, since
it assumes an elastic deformation of the spin texture that is uniform across the sample.  On the other hand, in the real material with disorder, it pays off to put the bulk of the deformation across weak links.  Since the helical spin correlations in \nigas have wave vector near ${\bf Q}=(1/6,1/6)$, Eq.~(\ref{eq:helicalRho}) gives the upper bound,
\begin{eqnarray}
\rho_{zz}^{max}=2\rho_{xx}^{max} =2\rho_{yy}^{max}\approx 4\sqrt{3} J_3 \langle S_i \rangle^2/d,\label{eq:helRho}
\end{eqnarray}
Using $|\Theta_{CW}|=80$ K and the ratio\cite{nakatsuji1} $J_1/J_3=-0.2$, we obtain $J_3\approx k_B (25 {\rm K})$.
Plugging in the experimental values \cite{nakatsuji1}, $g=2$, $|\langle {\bf S}_i\rangle|\approx 0.75$, we obtain an upper bound,
$\rho_{xx}^{max}d = k_B (49\,{\rm K})$, which is indeed larger than $E_2$.  Conversely, we get a lower bound 
$c_M/T^2>2.9\times 10^{-3} {\rm J\, mole^{-1}K^{-3}}$
that is consistent with the experimentally measured value, which is about 9 times larger than this bound.   In fact, the ratio between
the two is comparable to the frustration parameter, $T_f / |\Theta_{CW}|$,
which is a natural renormalization factor for the spin stiffness in a frustrated spin system.

Note that the definition of the two energy scales $E_1$ and $E_2$ is {\it model independent}, and can therefore be used to characterize the experiments and to compare with other theoretical proposals.  As we have argued, in the HS scenario, it is natural to find $E_2\sim T_f \ll E_1\sim |\Theta_{CW}|$.  On the other hand, for the spin nematic states of Refs.~\cite{tsunetsugu} and \cite{senthil}, this requires some fine tuning.  For example, in order to obtain $E_2\ll |\Theta_{\rm CW}|$, the system has to be tuned to be very close to a quantum critical point.

{\it Effect of magnetic field and spin anisotropy.}-- 
One of the salient features of \nigas is the weak effect of magnetic fields on the specific heat anomaly.  For a typical magnetically-ordered system, an applied field gaps out some of the collective modes and
therefore suppresses the specific heat below a temperature of order $T^*=g\mu_B H/k_B$. 

\begin{figure}
\includegraphics[width=3.2 in]{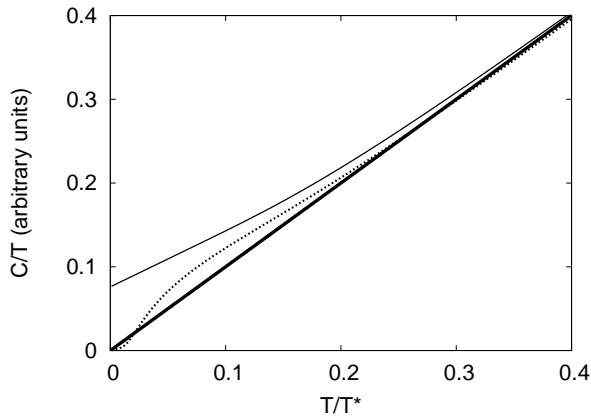}
\caption{Specific heat for an isotropic system (thick solid line); for an isotropic system in a magnetic field (dashed line); and for a system with easy-plane anisotropy with anisotropy energy $0.3 T^*$, together with an magnetic field perpendicular to the anisotropy plane (thin solid line).  Here, $T^*=g\mu_B H$.
\label{fig:Bdep} } \vskip-0.2in
\end{figure}

In the HS scenario, an applied field splits the 3 polarizations,
\begin{eqnarray}
\omega_0&=&v k \nonumber\\
\omega_\pm^2&=&(vk)^2+\frac{(g\mu_B H)^2}{2}\pm \sqrt{\frac{(g\mu_BH)^4}{4}+(g \mu_BHvk)^2}. \nonumber
\end{eqnarray}
Hence, $\omega_0$ is unchanged by the magnetic field and $\omega_+\approx g\mu_B H + \frac{v^2k^2}{g\mu_B H}$ is gapped.  However, the third mode, $\omega_-\approx  \frac{v^2k^2}{g\mu_B H}$,
obtains a quadratic spectrum at low energies.  This soft mode compensates for the gapped mode $\omega_+$, so that there is negligible deviation from the $T^2$ anomaly down to $T\approx T^*/4$
as can be seen in Fig.\ref{fig:Bdep}.  
This is consistent with the weak magnetic field dependence of the specific heat down to
the lowest temperature in the experiments.
Below $T\approx T^*/4$, $\omega_-$ dominates, leading to a low temperature regime with at $T$-linear specific heat anomaly. 
Notice that nuclear quadrupolar resonance (NQR) at Ga sites shows the existence of 
inhomogeneous internal field. The resulting distribution of Zeeman energy would further
mask the magnetic field dependence of the specific heat.

In realistic systems, anisotropy in the spin interactions will cut off the $T$-linear specific heat.   
Susceptibility measurements suggest that \nigas has a weak easy plane anisotropy.  In the absence of a magnetic field, this type of anisotropy introduces a gap in two of the three HS modes.    This gap is of order of the anisotropy energy, and is likely very small in \nigasn.

{\it Nuclear Magnetic Resonance.}--
NMR measurements on $^{69,71}$Ga nuclei yield a spin lattice relaxation $1/T_1\propto T^3$ at temperatures below 1 K  \cite{takeya}.  This is consistent with
the HS scenario.  The dissipative part of the dynamic spin susceptibility due to the HS modes is
\begin{eqnarray}
\chi''({\bf k},\omega)&=&\frac{\chi_0 D k^2 \omega}{2}\left[\frac{1}{(\omega-v k)^2+(Dk^2)^2}\right.\nonumber\\
&+&\left. \frac{1}{(\omega+vk)^2+(Dk^2)^2}\right]\nonumber
\end{eqnarray}
plus regular terms.  For modes propagating in two dimensions, this leads to $1/T_1\propto T^3$.  Note that, even if the HS modes are gapped due to anisotropy, two magnon Raman
scattering also yields $1/T_1\propto T^3$ \cite{moriya}.  On the other hand, for a nematic state, the dynamic structure factor acquires one more power of $|{\bf k}|$ at long wave lengths.
Thus, for a nematic state we expect $1/T_1$ to vanish at least as $T^4$ at low temperatures, in contrast to experiment.

{\it Direct detection of HS modes via inelastic neutron scattering and future experiments.}--
The HS modes are spin-one excitations, and therefore couple directly to neutrons.   Recent inelastic neutron scattering measurements indeed see
evidence of a linearly-dispersing mode centered at wave vector ${\bf Q}$, and with spin-wave velocity $v_{\rm neutron}=29$ meV \AA \cite{stock}.   On the other hand,
the spin-wave velocity extracted from the specific heat anomaly via Eq.\ref{specific_heat}
is much smaller,
\begin{eqnarray}
v_{\rm sh}=\sqrt{n_p} 5.4\,{\rm meV \AA}=9.35\,{\rm meV \AA},
\end{eqnarray}
where the last value corresponds to $n_p=3$ collective mode polarizations.   

This large difference between $v_{\rm sh}$ and $v_{\rm neutron}$ is due to 
the fact that $v_{\rm neutron}$ probes
the velocity of relatively high-energy spin-waves.  In the current measurements,  the inelastic neutron peaks can only be resolved clearly for
energies larger than 1 meV.  These relatively high-energy modes do not contribute directly to the specific heat anomaly below $T_f=8 K$.  In fact, at these 
energies the spin waves are expected to probe the {\it bare} stiffness.  An estimate for the bare stiffness tensor is the stiffness computed for the helical state, 
as given in Eq.~(\ref{eq:helRho}).  This yields two different spin-wave velocities,
\begin{eqnarray}
v_{x,\rm helical}=v_{y,\rm helical}=24\,{\rm meV \AA}\\
v_{z,\rm helical}=34\,{\rm meV \AA}.
\end{eqnarray}
These velocities are consistent with current experiments -- note that the peaks measured 
with neutrons are very broad, and thus it is very difficult to identify
two separate but nearby peaks.  

Future experiments may be able to distinguish the two separate spin-wave velocities.  In addition, with increased resolution it may be possible to measure
the dispersion of spin-waves to lower energies.  In the HS scenario, this would yield a spin-wave velocity that changes with wave vector, from $v_{\rm sh}$ very
close to ${\bf Q}$, to $v_{\rm helical}$ at wave vectors far from ${\bf Q}$. Hence, this would constitute a direct measurement of the length-scale dependent spin stiffness.
One further signature of the HS modes that may be seen in future neutron scattering experiments is the spin-wave damping, which is predicted to be \cite{HalperinSaslow}
\begin{eqnarray}
\Lambda=D ({\bf k}-{\bf Q})^2
\end{eqnarray}
We note that the low energy director modes that occur in the nematic states of Refs.~\cite{senthil,tsunetsugu} do not couple directly to neutrons at long wave lengths \cite{mila}, and therefore cannot account
for these inelastic neutron scattering peaks.

{\it Other considerations: two level systems.}--
In conventional spin-glasses, the $T^2$ contribution to specific heat coming from Halperin-Saslow modes is overwhelmed by a linear $T$ contribution
coming from localized two-level-systems (TLS)\cite{AndersonHalperinVarma},
\begin{eqnarray}
c_{tls}=\frac{\pi^2}{6}N_0 k_B^2 T.
\end{eqnarray}
Here, $N_0$ is the density of states of TLS.  A rough estimate for $N_0$ can be obtained by assuming that  the energy distribution
function of a single TLS is $1/(k_B T_f)$, and that there is one TLS in each correlation volume $v_\xi\sim \xi_{xy}^2 d$.  Using the in-plane correlation length $\xi_{xy}\sim7 a$, one obtains an approximate upper limit
$N_0 \lesssim 1/\xi_{xy}^2\xi_z k_B T_f$, leading to
\begin{eqnarray}
\frac{c_{tls}}{T}\lesssim 3.0\times 10^{-2} {\rm J\, mol^{-1} K^{-2}}.\label{eq:tls}
\end{eqnarray}
If present, such a contribution would be seen only at the lowest temperatures 
explored in Ref.~\cite{nakatsuji1}.  This suggests that the HS modes give the
dominant contributions in the temperature range studied in Ref.~\cite{nakatsuji1}.

{\it Conclusion.}--
We have shown that the phenomenology of \nigas below the freezing temperature $T_f$
can naturally be explained by the Halperin-Saslow scenario.  In this theory, linearly dispersing
HS modes in the background of frozen spin moments can exist for a much longer 
length scale than that of the short-range magnetic order.
This explains the simultaneous onset of the short-range helical magnetic order and
$T^2$ specific heat at $T_f$. The HS modes are consistent with $T^3$ spin-lattice relaxation
rate observed in the NMR experiments. 
We also showed that the specific heat anomaly in the HS scenario is only
weakly sensitive to applied magnetic field as observed in the experiments.
In more recent experiments on purer samples  \cite{MaenoNakatsujiPrivate}, 
indications of freezing start to show up at 10K. However, only at 2 K do the 
magnetic moments seem to be completely frozen  \cite{MaenoNakatsujiPrivate}.  
In the temperature range between 2 and 10 K, the moments may not be completely static.   
Given that the freezing temperature may depend on the time scale of the probe,
there might be some ambiguity associated with it, leading to some uncertainty of the
value of the spin stiffness in our analysis. Notice, however, that
the value of the experimentally-extracted spin stiffness $E_2=6.5$ K is in rough 
agreement with both extremes of the temperature range 2-10 K.

It was pointed out that the HS modes may have already been observed in 
inelastic neutron scattering at relatively high energies,
although current experimental resolution does not give access to the low 
energy mode dispersion relation. We argued that 
the spin-wave velocity measured at relatively high energies reflects the 
bare spin stiffness and the velocity of the low energy HS modes, that is
directly related to the low temperature specific heat, should be
different at lower energy scales. We provided self-consistent 
estimates of both the low and high energy spin wave velocities.
These predictions as well as the three
polarizations of the spin waves and the detailed magnetic field dependence 
of the HS modes can be tested in future neutron scattering experiments on
single crystals.

We thank Y. Maeno, S. Nakatsuji, T. Senthil, and C. Stock for helpful discussions.
This work was supported by the NSERC of Canada, Canadian Institute for Advanced Research,  
Canada Research Chair, and KRF-2005-070-C00044.

\end{document}